\documentclass[10pt,conference]{IEEEtran}
\usepackage[english]{babel}
\usepackage{listings}
\usepackage{color}
\usepackage{colortbl}
\usepackage{comment}
\usepackage{graphicx}
\usepackage{epsfig}
\usepackage{array, colortbl}
\newcolumntype{P}[1]{>{\raggedright\arraybackslash}p{#1}}
\usepackage{listings}
\usepackage{epstopdf}
\usepackage{multirow}
\usepackage{rotating}
\usepackage{subfig}
\usepackage{float}
\usepackage[obeyspaces,hyphens,spaces]{url}
\usepackage{balance}
\usepackage{fancybox}
\usepackage{scalefnt}
\usepackage[normalem]{ulem}
\usepackage{cleveref}
\usepackage{siunitx}
\usepackage[hang]{footmisc}
\usepackage{booktabs}
\usepackage{pbox}
\usepackage{ragged2e}
\usepackage{fancybox}
\usepackage{guehene}

\newenvironment{myindentpar}[1]%
{\begin{list}{}%
         {\setlength{\leftmargin}{#1}}%
         \item[]%
}
{\end{list}}

\newcommand{\BLUE}[1]{{\color{black}#1}}

\makeatletter
\renewcommand{\paragraph}[1]{\noindent\textsf{#1}.}

\title{Is It Safe to Uplift This Patch? \\ \LARGE{An Empirical Study on Mozilla Firefox}}

\author{
	Marco Castelluccio \\
	Mozilla Corporation, United Kingdom \\
	DIETI, Universit\`a Federico II, Italy \\
	mcastelluccio@mozilla.com
	\and
	Le An and Foutse Khomh \\
	SWAT Lab \\
	Polytechnique Montr\'{e}al, QC, Canada\\
	\{le.an, foutse.khomh\}@polymtl.ca \\

}

\begin{document}
\maketitle

\begin{abstract}
In rapid release development processes, patches that fix critical issues, or implement high-value features are often promoted directly from the development channel to a stabilization channel, potentially skipping one or more stabilization channels. This practice is called \emph{patch uplift}. Patch uplift is risky, because patches that are rushed through the stabilization phase can end up introducing regressions in the code.
This paper examines patch uplift operations at Mozilla, with the aim to identify the characteristics of uplifted patches that introduce regressions. Through statistical and manual analyses, we quantitatively and qualitatively investigate the reasons behind patch uplift decisions and the characteristics of uplifted patches that introduced regressions. Additionally, we interviewed three Mozilla release managers to understand organizational factors that affect patch uplift decisions and outcomes. Results show that most patches are uplifted because of a wrong functionality or a crash. Uplifted patches that lead to faults tend to \BLUE{have larger patch size}, 
and most of the faults are due to semantic or memory errors in the patches. Also, release managers are more inclined to accept patch uplift requests that concern certain specific components, and--or that are submitted by certain specific developers. 

\end{abstract}


\IEEEpeerreviewmaketitle

\section{Introduction}
\label{sec:intro}
\newcommand{\RQone}{What are the characteristics of patches that are uplifted?}
\newcommand{\RQtwo}{What are the characteristics of uplifted patches that introduced faults in the system?}
\newcommand{\RQthree}{Can we improve the patch uplift process?}

The advent of continuous delivery and rapid release practices have significantly reduced the amount of stabilization time available for new features, forcing companies to resort to innovative techniques to ensure that important features are released to the public, in a timely manner and with a good quality. To cope with short release cycles, Mozilla has re-organized its release process around four channels: a development channel named \emph{Nightly}, two stabilization channels (\emph{Aurora} and \emph{Beta}), and a main \emph{Release} channel. Features corresponding to a new release are developed on the Nightly channel over a period of six weeks. After that, the code is transferred to Aurora, where it is tested by Mozilla developers and contributors, for a period of six weeks, and then to Beta where it is tested by a selected group of external users. Finally, mature Beta features are imported into the main Release channel and delivered to end users. This pipelined process allows Mozilla to avoid mixing the development of new features with the stabilization process, which is particularly important given that integration operations are unpredictable~\cite{Laforge}, and can significantly delay a release process, if not enough time is allowed for stabilization. 
However, this well organized release process is frequently subverted by urgent patches, implementing high-value features or critical fixes, that cannot wait for the next release train. These features and fixes are directly promoted from the development channel to stable channels (\ie{} Aurora, Beta, and main Release), a practice called \emph{patch uplift}. Patch uplift is risky because the time allowed for the stabilization of uplifted patches is reduced by six weeks for each skipped channel. Therefore, it is important to carefully pick the patches that are uplifted and ensure that developers scrutinize them properly, to reduce the risk of regressions. 
There are a set of rules in place at Mozilla to govern this uplift process. 
One of these rules is that patches uplifted to the Beta channel should be \emph{(1) ideally reproducible by the QA team, so that they can be verified; (2) should have been verified on Aurora/Nightly first; and (3) should not contain string changes (\ie{} changes in the text which is visible to users)}. However, despite these rules, multiple uplifted patches still introduce regressions in the code. Hence, it is unclear if--and--how the rules are enforced at Mozilla and why certain uplifted patches introduce post-release bugs.

%
In this paper, we conduct a series of quantitative and qualitative analyses
to understand the decision making process of patch uplift at Mozilla and the characteristics of uplifted patches that introduce regressions. 
Overall, we analyze 33,664 issue reports (corresponding to 7,267 uplift requests) in 17 versions of Firefox over a period of two years and answer the following research questions:

\begin{description}
\item[\textit{RQ1:}] \textit{\RQone}
\end{description}
\begin{myindentpar}{0.5cm}
We observe that most patches are uplifted to resolve wrong functionalities or crashes. Rejected uplift requests required longer decision time than accepted requests. We attribute this difference to the high complexity of these rejected patches (since complex patches require longer time for risk assessment). Last but not least, 
release managers tend to trust patches that concern certain specific components, and--or that are submitted by certain specific developers.
\end{myindentpar}

\begin{description}
\item[\textit{RQ2:}] \textit{\RQtwo}
\end{description}
\begin{myindentpar}{0.5cm}
From our analysis, we observe that uplifted patches that lead to faults tend to \BLUE{have larger patch size}; 
suggesting that developers and release managers need to carefully review patch candidates for uplift with a large amount of changes, before allowing for their uplift. Most faulty uplifts are due to semantic or memory-related errors. We also observed that patches related to certain components and--or submitted by certain developers are more likely to cause faults.
\end{myindentpar}



\textbf{The remainder of this paper is organized as follows.}
Section \ref{sec:back} provides background information about patch uplift. Section \ref{sec:design} describes the design of our case study. Section \ref{sec:case_study_results} presents the results of the case study, and Section~\ref{sec:discussion} elaborates on the implications of these results. 
Section~\ref{sec:threats} discusses threats to the validity of this study. Section~\ref{sec:related_work} summarizes related works, and Section~\ref{sec:conclusion} concludes the paper.

\section{Mozilla Patch Uplift Process}\label{sec:back}
This section describes the Mozilla patch uplift process and the rules governing this process.

Firefox follows a pipelined release process \cite{khomh2012faster}, with four release channels (\textit{Nightly}, \textit{Aurora}, \textit{Beta}, and \textit{Release}). New feature work is done on the \textit{Nightly} channel, while \textit{Aurora} and \textit{Beta} serve as stabilization channels, and the \textit{Release} channel is used to deliver the software to end users.
Every six weeks, there is a \emph{merge day}, when the code from a less stable channel flows into a more stable one (\eg{} the Nightly code is moved in the Aurora repository). 
Most of the development work is performed in the \textit{Nightly} channel, where patches can be committed after a normal review process. For the stabilization channels, a different process for committing patches has been put in place (\ie{} patch uplift), to keep the channels as stable as possible (as code committed to Aurora and Beta is closer to be released to users). Patches with important features or severe fault fixes that cannot wait for the entire process are promoted directly from the development channel to one of the stable channels, skipping the stabilization phase on one or more channels. 



The lifecycle of an uplifted patch can be summarized as follows: 
developers write a patch, which gets reviewed by one or more reviewers. After a successful review, the patch is committed to the Nightly channel. If developers (or other stakeholders) believe that the patch is particularly important (\eg{} it fixes a frequent crash, or a performance issue), they can ask for approval to uplift the patch to one (or more) of the stable channels, \ie{} Aurora, Beta, or Release.

Release managers (who are independent and different from reviewers) are responsible for deciding which patches can be uplifted. They can either \emph{accept} or \emph{reject} the patch uplift request, after a careful consideration of the risks involved.


The more a channel is stable, the higher is the bar for approval of uplift requests. Below we present an excerpt of the rules in place at Mozilla on the different channels.\\ 
\emph{Aurora}: Uplifts to the Aurora channel are less critical, as they still have considerable time for stabilization. The rules are not strict in this case: no new features are accepted; no disruptive refactorings; no massive code changes; no string changes, unless the localization team is aware and has approved; they must be accompanied, if possible, by automated tests.\\
\emph{Beta}: Uplifts to the Beta channel are more critical, as they have less time for stabilization. In addition to the rules outlined for Aurora, the changes uplifted to the Beta channel should be (1) ideally reproducible by QA, so that they can be verified; (2) they should have been verified on Aurora/Nightly first; and should not contain (3) changes to the user-visible strings in the application (as those require a very high effort and time to be localized, since Mozilla relies on volunteer contributors). The uplifted changes can be proven performance improvements, fixes to important crashes, fixes for recent regressions. The closer to the release date, the stricter the release managers should be in enforcing the rules.\\
\emph{Release}: Uplifts to the Release channel are generally discouraged, as they require a new version to be built and released to users. Possible uplifts are fixes for major top crashes, security issues, functional regressions with a very broad impact.

Once a patch is accepted for uplift, Tree Sheriffs \cite{TreeSheriffs} (\ie{} engineers responsible for supporting developers in committing patches and ensuring that the automated tests are not broken after commits, monitoring intermittent failures and backing out patches in case of test failures) or the developers themselves can commit it to the stabilization channel(s) for which the patch was approved. 
\section{Case Study Design}\label{sec:design}

In this section, we describe the data collection and analysis approaches that we use to answer our two research questions.

\subsection{Data Collection}\label{data_collection}
We collect, from the Mozilla issue tracking system (Bugzilla), all issues marked as \emph{resolved} or \emph{verified} in the Firefox and Core products between July 2014 (release date of Firefox 31.0) and August 2016 (release date of Firefox 48.0). In total, there are 35,826 issue reports in our dataset.

\begin{figure}[t]
\centering
\includegraphics[width = \linewidth]{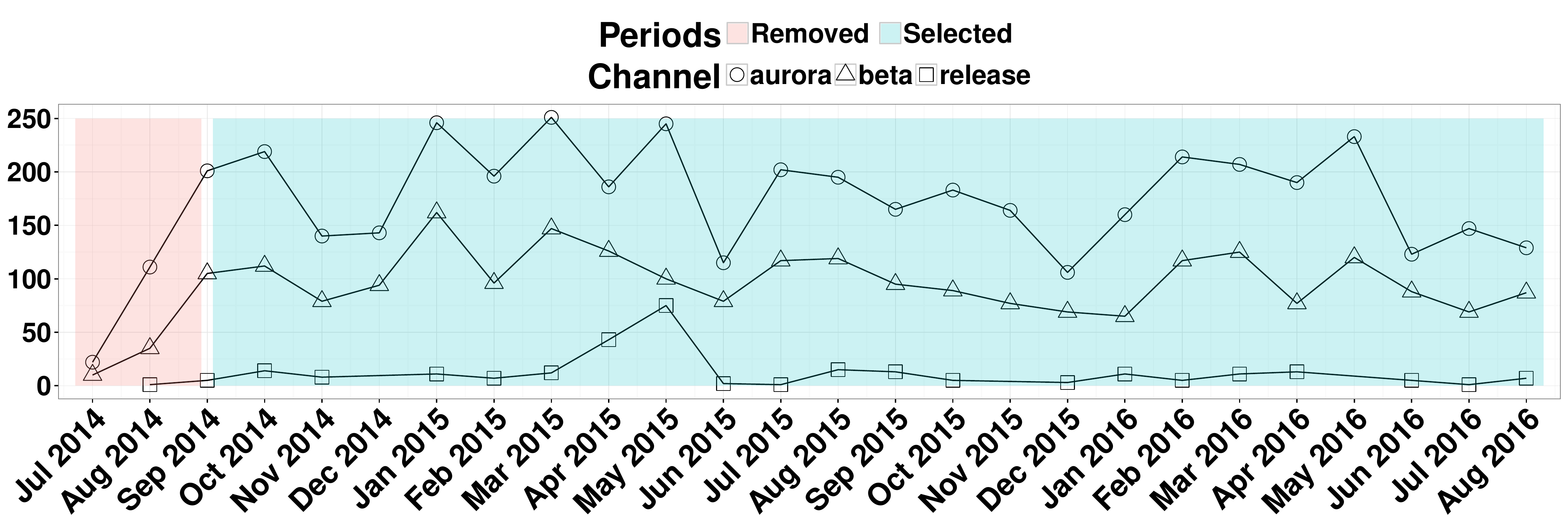}
\caption{Number of uplifts during each month from July 2014 to August 2016. Periods with low number of uplifts or not covering all the three channels are removed.}
\label{fig:study_period}
\end{figure}

Mozilla developers use customized Bugzilla flags to request for patch uplifts. These flags have the form \texttt{approval-mozilla-CHANNEL}, where \texttt{CHANNEL} can be Aurora, Beta, or Release. The postfix of the flag is set to a question mark (\texttt{?}) when a developer asks for an uplift, to a minus sign (\texttt{-}) if the release manager rejects the uplift, and to a plus sign (\texttt{+}) if the release manager approves the uplift. We rely on these flags to identify uplifted patches. At Mozilla, release managers usually inspect all patches in an issue report before deciding whether they can be uplifted together. Thus, in this work, we consider uplift characteristics at the issue level. If an issue contains multiple patches, we bundle the patches together. 
To study the patch uplift process, we need to consider a period of time during which the practice was well established at Mozilla. To decide on this period, we computed the amount of patches that were uplifted each month, over our initial period of July 2014 to August 2016. Figure~\ref{fig:study_period} shows the distribution of the number of uplifts in three Firefox's release channels during this period. We do not consider uplifts that concern the ``Pocket'' component, as the inclusion of Pocket (which is a third-party add-on) in Firefox, a one-time event, might introduce noise in our data.
In Figure~\ref{fig:study_period}, each time point represents a period of one month (we can see that the Release channel did not receive any uplift in May and November 2015). 
Figure~\ref{fig:study_period} shows that the number of uplifted patches increased from July 2014 to August 2014 and then became stable from September 2014 to August 2016. Based on this distribution, we selected the period between September 2014 and August 2016, for our study. In other words, we limited our dataset to only issue reports and commits that occurred within this period. Between September 2014 and August 2016, we study in total 33,664 issue reports, 
in which there are 7,267 uplift requests: 285 to Release, 2,614 to Beta, and 4,368 to Aurora.

\subsection{Data processing}
%
\begin{figure}[t]
\centering
\includegraphics[width = \columnwidth]{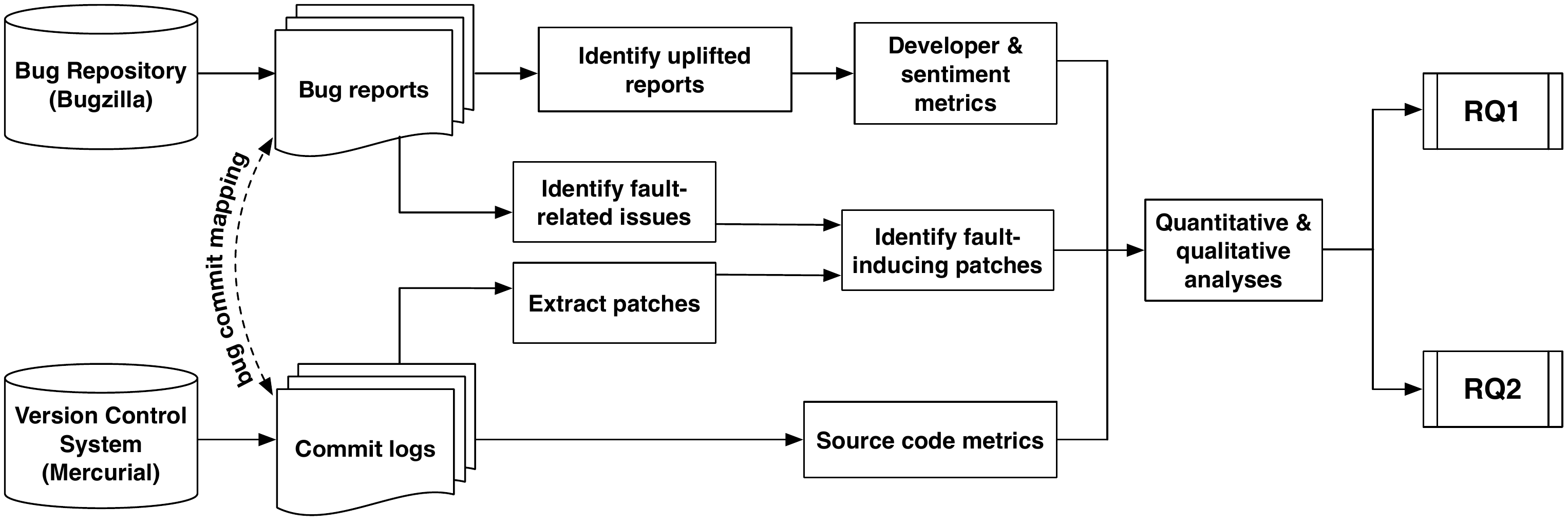}
\caption{Overview of our data processing approach.}
\label{fig:flow_graph}
\vspace{-10pt}
\end{figure}
Figure~\ref{fig:flow_graph} shows a general overview of our approach. We describe each step of the approach below. The corresponding data and scripts are available online at: \url{https://github.com/swatlab/uplift-analysis}.

\subsubsection{Identification of Fault-related Issues}
\label{sec:identificationFaultRelatedIssues}
Mozilla uses Bugzilla to manage and track its issues. All types of issues, whether they are faults or new features, are managed in this system. Unlike JIRA~\cite{jira}, which offers the possibility to distinguish between issues using a tag, Bugzilla does not provide issue type information. Therefore, our first processing task is to differentiate issues that are related to faults, from new feature requests or improvements. To automatically identify fault-related issues, we use a keyword-based heuristic to search information in the title, description, \BLUE{flags}, and user comments of each issue report. Our list of keywords includes: crash, regression, failure, leak, steps to reproduce (STR), and hang. \BLUE{The full list is available at: \url{https://github.com/swatlab/uplift-analysis}}.

To ensure the accuracy of our detection on fault-related issues, we manually validated a sample of our results. From a total of 33,664 issue reports, we randomly selected a sample of 380 issue reports, which corresponds to a confidence level of 95\% and a confidence interval of 5\%. The first and the second authors read each of the 380 issue reports independently and classified them into \emph{fault-related} and \emph{other} categories. We then compared their classification results and observed that 41 issue reports were classified into different categories by the two authors. To resolve these discrepancies, we created an online document for the 41 issues; allowing all of the authors to comment and discuss the issues. After this round, a consensus was reached for 35 out of the 41 issues. For the remaining 6 issues, we organized a meeting and discussed the classification of each of them until a consensus was found. 
The result of our manual classification shows that our keyword-based heuristic achieves a precision of 87.3\% and a recall of 78.2\%, when classifying issues into \emph{fault-related} and \emph{other} categories.

\subsubsection{Identification of Fault-inducing Patches}\label{sec:szz}
We use the SZZ algorithm~\cite{sliwerski2005changes} to identify patches (these patches could be fault-fixing patches or patches related to features or improvements) that introduced faults in the system. First, we used Fischer et al.'s heuristics~\cite{fischer2003populating} to map each studied issue to its corresponding patch(es) (\ie{} commits). This heuristic consists in looking for issue IDs in commit messages using regular expressions. 
Next, for each fault-related issue, we use the following Mercurial command to extract the list of files that were changed to fix the issue:\\ 
\texttt{\footnotesize{hg log --template \{commit\},\{file\_mods\},\{file\_dels\}}}

\noindent
In this step, we only consider modified and deleted lines, since added lines could not have been changed by prior commits. We denote an issue's fault-fixing file by $F_{fix}$. Then, for each changed file $f_{fix} \mid f_{fix} \in F_{fix}$, we use Mercurial's \texttt{annotate} command as follow to check which prior commits changed the lines that were modified by the fault-fixing commits. The SZZ algorithm assumes that the fault is located in these lines.\\ 
\texttt{\footnotesize{hg annotate commit\^{} -r f\_{fix} -c -l -w -b -B}}

\noindent
We refer to the obtained commits as \emph{fault-inducing candidates}. Finally, we examine whether a fault-inducing candidate was submitted before the creation date of its corresponding fault-related issue report. If so, we consider the candidate to be a \emph{fault-inducing commit}, and its related issue to be a \emph{fault-inducing issue}.

\subsubsection{Mining Issue Reports}
We mine several kinds of metrics from Bugzilla issue reports: information about the review process (\eg{} how long a review took, how many reviewers inspected a patch), information about the uplift process (\eg{} whether an uplift was accepted, how long before a release manager decided to accept or reject an uplift request), the developer assigned to an issue, and the component(s) affected by an issue. 

\subsubsection{Computing Metrics}\label{sec:all_metrics}
To capture the characteristics of patches that were uplifted, we computed the 22 metrics described in Table \ref{tab:all_metrics}. These metrics correspond to the following five dimensions: 

\begin{table}[t]
	\centering
	\scriptsize
	\caption{Metrics used to compare patches.}
	\begin{tabular}{ | >{\RaggedRight} p{1.6cm} | c | p{5.5cm} | }
		\hline
		\textbf{Metric} & $m_i$ & \textbf{Description} \\ \hline
\multicolumn{3}{|c|}{\textbf{Developer experience and participation metrics ($m_{1}$ - $m_{5}$)}}\\ \hline		
		Developer experience & 1 & Number of previous commits of the patch developer.\\ \hline
		Reviewer experience & 2 &  Number of previous commits of the patch reviewer.\\ \hline
		Number of comments & 3 & Number of comments in the issue report. \\ \hline
		Comment words & 4 & Average number of words in the comments to an issue.\\ \hline
		Review duration & 5 & Time period (in days) from a patch's submission until its approval.\\ \hline
\multicolumn{3}{|c|}{\textbf{Uplift process metrics ($m_{6}$ - $m_{8}$)}}\\ \hline
		Landing delta & 6 & Time elapsed (in days) between when the patch was applied to the Nightly version and when the developer asked for approval of an uplift. \\ \hline
		Response delta & 7 & Time elapsed (in days) between when the developer asked for approval for the uplift and when the release manager decided (approved or rejected). \\ \hline
		Release delta & 8 & Time elapsed (in days) between when the developer asked for approval for the uplift and the date of the following release. \\ \hline
\multicolumn{3}{|c|}{\textbf{Sentiment metrics ($m_{9}$ - $m_{10}$)}}\\ \hline
		Developer sentiment & 9 & The highest negative sentiment score in the developers' comments on an issue.\\ \hline
		Owner sentiment & 10 & The highest negative sentiment score in module owners' comments on an issue.\\ \hline
		\multicolumn{3}{|c|}{\textbf{Code complexity metrics ($m_{11}$ - $m_{19}$)}}\\ \hline
		Patch size & 11 & Number of lines in a patch (excluding test patches). \\ \hline
		Test patch size & 12 & Number of lines in a test patch.\\ \hline
		Prior changed times & 13 & Number of previous commits that modified the same files that the patch is modifying. \\ \hline
		LOC & 14 & Average lines of code in all classes in a patch.\\ \hline
		Average cyclomatic & 15 & Average cyclomatic complexity of the functions in a class.  \\ \hline
		Number of functions & 16 & Average number of classes' functions in a patch. \\ \hline
		Maximum nesting & 17 & Average maximum level of nested functions in all classes in a patch.\\ \hline
		Comment ration & 18 & Average ratio of the lines of comments over the total lines of code in all classes in a patch.\\ \hline
		Module number & 19 & Number of modules involved by a patch.\\ \hline
		\multicolumn{3}{|c|}{\textbf{Code centrality (SNA) metrics ($m_{20}$ - $m_{22}$)}}\\ \hline
		PageRank & 20 & Time fraction spent to ``visit'' a class in a random walk in the call graph.\\ \hline
		Betweenness & 21 & Number of classes passing through a class among all shortest paths. \\ \hline
 		Closeness & 22 & The average length of the shortest path between a class and all other classes. \\ \hline	
	\end{tabular}
	\label{tab:all_metrics}
\vspace{-15pt}
\end{table}

\paragraph{Developer experience and participation metrics}
Our rationale for computing these metrics is that patches written or reviewed by experienced developers may have a higher chance to be accepted for uplift, and may be less fault-prone. Long comments and long review durations may indicate the complexity of an issue and developers' uncertainty about it, which may explain its rejection or fault-proneness. 

\paragraph{Uplift process metrics}
We compute metrics capturing the uplift process for the following reasons. Release managers may be more inclined to accept patches with higher landing delta (as the more time a patch has been on the Nightly channel, the more time it has been tested by Nightly users). Patches with low release delta are likely to be refused uplifts, since patches that are developed closer to the date of release might pose more risk (as there is less time to fix potential regressions). Patches with low response delta may also be rejected (since developers have less time to evaluate the risks associated with the patch). 
Patches with low landing delta, release delta, and low response delta may also lead to faults if uplifted. 

\paragraph{Sentiments}\label{sec:complexity}
We compute sentiment metrics because we believe that sentiments can affect uplift decisions and their success rate. 
From each studied issue, we extract developers' comments to compute their sentiments. We leverage the sentiment mining tool, \emph{SentiStrength}~\cite{mike2010sentiment}, to estimate the extent of developers' positive and negative sentiments toward a specific issue. As one of the state-of-the-art sentiment mining tool, SentiStrength is easy to apply, and it has achieved a reasonable performance in prior works~\cite{mike2010sentiment}. 
In addition to developers' sentiments, we also computed module owners' sentiments. 

\paragraph{Code Complexity}\label{sec:complexity}
Previous works, such as \cite{kim2011crashes}, have shown that complex code is likely to introduce faults. We calculate code complexity metrics to understand how uplifting decisions and their success are affected by the complexity of the uplifted patches. 
We extract the files changed in each patch and use the static code analysis tool \emph{Understand}~\cite{understand_tool} to calculate the following complexity metrics on the files: lines of code (LOC), average cyclomatic complexity, number of functions, maximum nesting, and ratio of the comment lines over the total code lines. 

\paragraph{Code centrality (SNA) metrics}\label{sec:sna}
Kim et al.~\cite{kim2011crashes} observed that functions close to the centre of a call graph are likely to experience more faults. Hence, we compute metrics capturing the centrality of functions involved in uplifted patches and uplifted patch candidates. We use the network analysis tool, \emph{igraph}~\cite{csardi2006igraph}, in combination to \emph{Understand}~\cite{understand_tool}, as in~\cite{qrs2015}, to compute the following \emph{Social Network Analysis} (SNA) metrics: PageRank, betweenness, and closeness.
%
When computing complexity and SNA metrics, we only consider the C/C++ code since Firefox contains 86\% of C/C++ code.  
Computing code complexity and SNA metrics is a very time-consuming task. Instead of computing the metrics for each patch, we compute metrics by releases and map a given patch to its latest major release as in our previous work~\cite{qrs2015}. To make the metric results as precise as possible, we consider all major releases from Firefox 32.0 until Firefox 48.0, which cover the system's history from September 2014 until August 2016.



\section{Case Study Results}
\label{sec:case_study_results}

This section presents and discusses the results of our two research questions. For each question, we discuss the motivation, the approach designed to answer the question, and the findings.
To get a deeper insight of the patch uplift process, we perform both quantitative and qualitative analyses for each research question.


\subsection*{RQ1: \RQone}
\noindent
\emph{\textbf{Motivation.}}
This question aims to understand the characteristics of patches that are uplifted. We are particularly interested in understanding what differentiates patch uplifts among different channels. 
Although Mozilla has published rules to guide the patch uplift process~\cite{uplift_rules}, it is unclear if and how these rules are enforced in practice. 
The answer to this research question can help discover hidden factors that affect the uplift process, and help software practitioners make this process more predictable. 

\medbreak
\paragraph{1) Quantitative Analysis}

\emph{\textbf{Approach.}} Using the metrics from Table~\ref{tab:all_metrics}, we statistically compare 22 numerical characteristics of patch uplift candidates that were accepted and those that were rejected. 
As Mozilla release managers take a whole issue report into account during the uplift process (see Section \ref{data_collection}), we calculate the average values of the code complexity and SNA metrics for all patches in a subject issue report.

For each of the 22 metrics $m_i$, we formulate the following null hypothesis:\\
$H_i^{01}$: \emph{there is no difference between the values of $m_i$ for patch uplift candidates that were accepted and those that were rejected}, where $i \in \{1, \ldots, 22\}$

We use the Mann-Whitney U test~\cite{hollander2013nonparametric} to accept or reject these hypotheses. 
The Mann-Whitney U test is a non-parametric statistical test that measures whether two independent distributions have equally large values. 
We use a 95\% confidence level (\ie{} $\alpha = 0.05$) to accept or reject the hypotheses. 
Since we perform more than one comparison on the same dataset, to reduce the chances of obtaining false-positive results, we use Bonferroni correction~\cite{dmitrienko2005analysis} to control the familywise error rate. Concretely, we calculate the adjusted $p$-value, which is multiplied by the number of comparisons. Whenever we obtain 
statistically significant differences between metric values, we compute the Cliff's Delta effect size~\cite{cliff1993dominance} to measure the magnitude of the difference. Due to the page limit, we will only report the metrics for which there is a statistically significant difference between accepted and rejected patch uplift candidates.


\begin{table}[t]
\centering\scriptsize
 \caption{Accepted vs. rejected patch uplift candidates.}
 \label{tab:rq1_res}
 \begin{tabular}{lp{1.7cm}rrrr}
	\toprule
    Channel
     & Metric          & Accepted & Rejected & $p$-value & Effect size \\
    \midrule
    \emph{Aurora}
    & Comment ratio & 0.1 & 0.2 & \textbf{0.03} & small \\
    & Landing delta & 0.4 & 3.0 & \textbf{0.02} & small \\
    & Response delta & 0.9 & 2.4 & \textbf{1.80e-05} & medium \\
    \hline
    \emph{Beta}
    & LOC & 529.0 & 1,046.8 & \textbf{9.27e-04} & small \\
    & Cyclomatic & 2.0 & 3.0 & \textbf{0.04} & negligible \\
    & \# of functions & 20.0 & 35.2 & \textbf{9.62e-04} & small \\
    & Comment ratio & 0.1 & 0.2 & \textbf{8.86e-05} & small \\
    & Betweenness & 2,789.0 & 20,586.3 & \textbf{0.01} & negligible \\
    & PageRank & 1.4 & 1.7 & \textbf{0.01} & negligible \\
    & Max. nesting & 2.3 & 3.0 & \textbf{7.72e-03} & negligible \\
    & Module number & 1.0 & 1.0 & \textbf{7.13e-03} & negligible \\
    & Response delta & 0.7 & 1.0 & \textbf{6.28e-04} & small \\
    \hline
    \emph{Release}
    & Response delta & 0.02 & 3.1 & \textbf{1.39e-12} & large \\
    \bottomrule
 \end{tabular}
 \vspace{-15pt}
\end{table}

\smallbreak
\emph{\textbf{Results.}}
Table \ref{tab:rq1_res} summarizes differences between the characteristics of patches that were accepted for an uplift and those that were rejected. 
We show the median value of accepted and rejected uplifts for each metric, as well as the $p$-value of the Mann Whitney U test and the effect size. For all three channels, rejected uplifts have longer response delta \BLUE{($m_7$)} than accepted uplifts. We attribute this outcome to the high complexity of the rejected patches, which required longer time for risk assessment. We summarize the different results among the channels as follows:

\begin{itemize}
    \item \emph{Aurora}: We observe that rejected uplift requests have significantly higher landing delta; this might imply that the rejected patches are landing at the end of the Aurora cycle, and so have less time for stabilization. Also, rejected uplift requests have higher ratio of comment in the source code, although we expected that a higher comment ratio might help release managers understand the code. A high comment ratio could also indicate a high code complexity. Release managers may hesitate to release patches with complex code ahead of schedule.
    \item \emph{Beta}: Compared to accepted patches, rejected patches tend to have higher code complexity in terms of LOC and number of functions, as well as higher SNA values in terms of PageRank. This result is expected, because we assume that complex code and code connected with many other classes is less likely to be accepted for urgent releases. As in the Aurora channel, rejected patches also contain code with higher ratio of comment. Although accepted and rejected patches have significant differences on some other metrics such as cyclomatic complexity, the magnitude of these differences is negligible.
\end{itemize}

\textbf{According to the results, we can only reject $H_{7}^{01}$, meaning that the response delta can significantly affect the decision to uplift a patch or not. The impact of other metrics, including code complexity and SNA metrics, is channel dependent.}

We quantified the acceptance rate of uplift requests for different components and observed that 
certain components enjoy a 100\% acceptance rate (perhaps because they rarely experienced faults); while other components have lower acceptance rates (perhaps because they are inherently more complex, \eg{} the implementation of JavaScript, or because release managers have had bad experience with some of them). This difference between the acceptance rates of components is more pronounced in the Release channel. Some components that are involved in a large number of uplifts (\eg{} \textit{Audio/Video}, \textit{Graphics}, and \textit{DOM} components) also have the lowest acceptance rate. Perhaps developers of those components tend to ask for uplifts more often, prompting a negative reaction from release managers who may feel that they take too many risks.

\begin{table}[t]
	\centering
	\scriptsize
	\caption{Uplift reasons and descriptions (abbreviations are shown in parentheses).}
	\begin{tabular}{ | >{\RaggedRight} p{2cm} | p{6cm} | }
		\hline
		\textbf{Reason} & \textbf{Description} \\ \hline		
		Security & Security vulnerability exists in the code.\\ \hline
		Crash & Program unexpectedly stops running.\\ \hline
		Hang & Program keeps running but without response.\\ \hline
		Performance degradation (perf) & Functionalities are correct but response is slow or delayed.\\  \hline
		Incorrect rendering (rendering) & Components or video cannot be correctly rendered.\\ \hline
		Wrong functionality (func) & Incorrect functionalities besides rendering issues. \\  \hline
		Web incompatibility (web comp) & Program does not work correctly for a major website or many websites due to incompatible APIs or libraries, or a functionality, which was removed on purpose, but is still used in the wild. \\ \hline
		Add-on or plug-in incompatibility (addon comp) & Program does not work correctly for a major add-on/plug-in or many add-ons/plug-ins due to incompatible APIs or libraries, or a functionality, which was removed on purpose, but is still used in the wild. \\ \hline
		Compile & Compiling errors. \\ \hline
		Feature & Introduce or remove features, including support adding. \\ \hline
		Improvement (improve) & Minor functional or aesthetical improvement. \\ \hline
		Test-only problem (test) & Errors that only break tests.\\ \hline
		Other & Other uplift reasons, \eg{} data corruption and license incompatibility. \\ \hline
	\end{tabular}
	\label{tab:uplift_reasons}
\vspace{-15pt}
\end{table}


\medbreak
\paragraph{2) Qualitative Analysis}
Since we did not observed significant structural differences between the code of patch uplift candidates that were rejected and those that were accepted, we conducted a qualitative study to identify and compare the reasons behind successful and failed patch uplift requests. 

\emph{\textbf{Approach.}} 
From 2,384 uplifted issues in the Beta channel and 231 uplifted issues in the Release channel, we randomly choose respectively 459 and 154 issues as our samples (which correspond to a confidence level of 95\% and a confidence interval of 5\%). Inspired by Tan et al.'s work~\cite{tan2014bug}, we classify the uplift reasons into 14 categories based on their (potential) impact and detected fault types. Some of Tan et al.'s categories are too broad, such as incorrect functionality. We break them into more detailed uplift reasons, \eg{} incorrect functionality is split to incorrect rendering and (other) wrong functionality. Some of Tan et al.'s categories, such as data corruption, are with too few occurrences. We combine them into the ``other'' category.
Table \ref{tab:uplift_reasons} shows the uplift reasons used in our classification. We perform a card sorting on each of the sampled issues. By studying the issue report, the first and the second authors of the paper individually classified each issue into one or multiple uplift reasons (some uplift may be due to multiple reasons). Then we compared their classifications and resolved conflicts through discussions. 
We discussed each conflict until an agreement was reached.

To connect uplift reasons with the risk of regression, we will show the distribution of the faulty uplifts for each uplift reason. 

Moreover, to identify organization factors that play a role in patch uplift decisions, 
we interviewed three of the current five Mozilla release managers (the other remaining two are new to the role) \BLUE{one at a time (to avoid them influencing each other)}, asking them the following questions:
\begin{enumerate}
	\item \emph{Which factors do you take into account when deciding about an uplift?}
	\item \emph{Are there differences in how you handle uplifts in different channels, and what are the differences?}
	\item \emph{How do you decide which developers you can trust?}
\end{enumerate}

We also reported the results of our quantitative analysis to them and asked for their feedback.

\begin{figure}[t]
\centering
\includegraphics[width = \columnwidth]{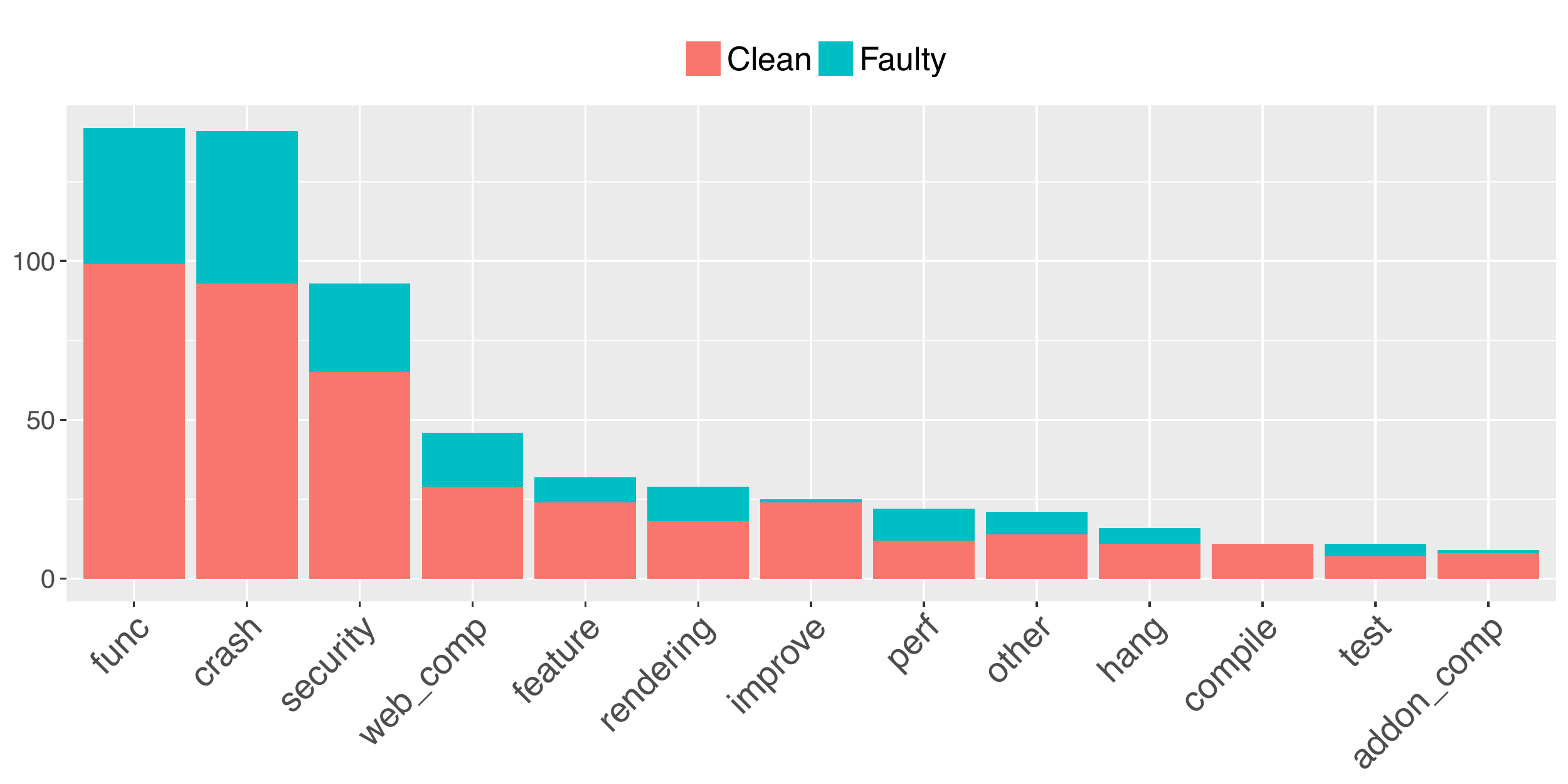}
\caption{Distribution of uplift reasons in Beta.}
\label{fig:uplift_beta}
\vspace{-10pt}
\end{figure}

\begin{figure}[!]
\centering
\includegraphics[width = \columnwidth]{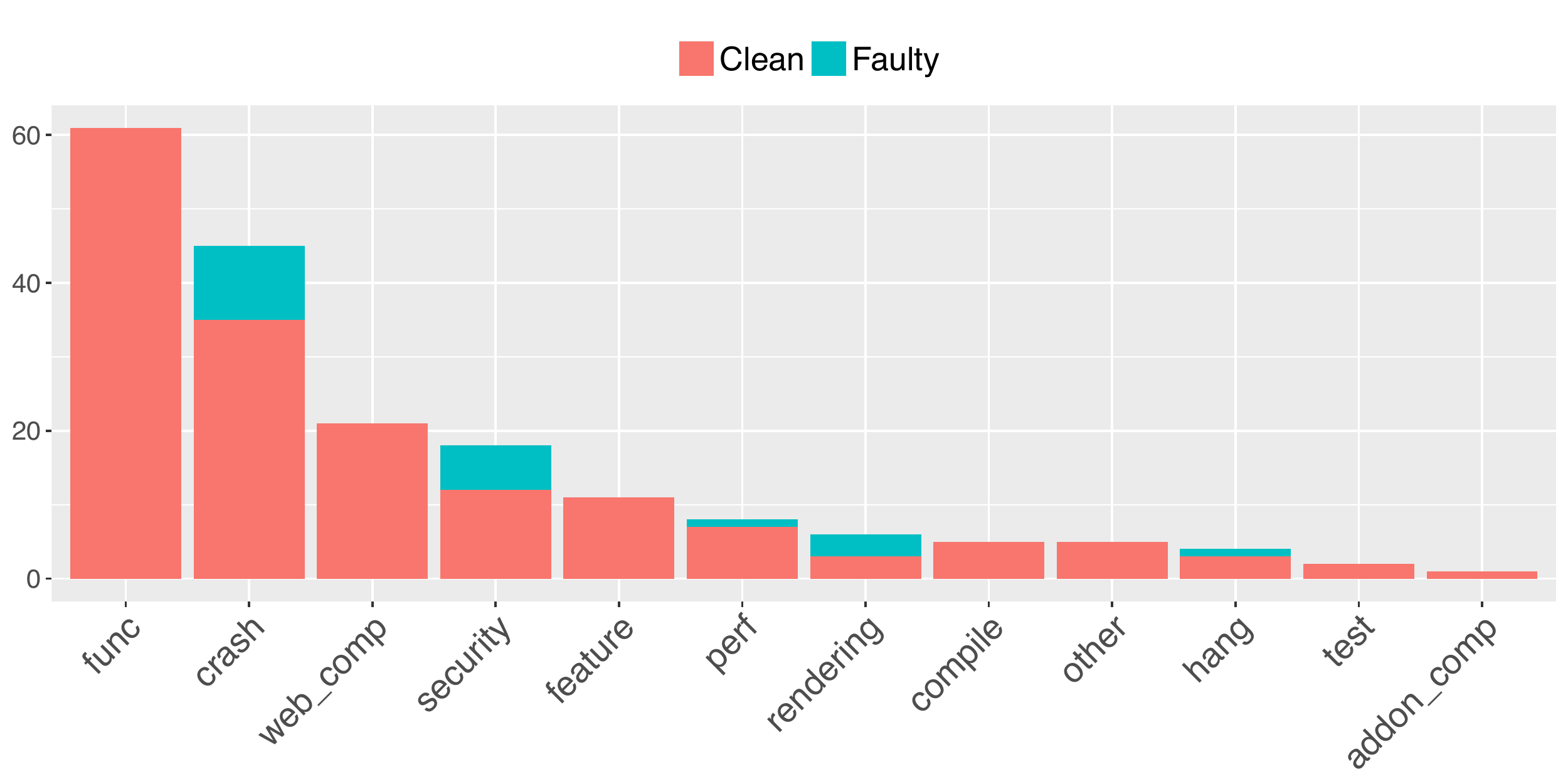}
\caption{Distribution of uplift reasons in Release.}
\label{fig:uplift_release}
\vspace{-10pt}
\end{figure}

\smallbreak
\emph{\textbf{Results.}} Figures \ref{fig:uplift_beta} and \ref{fig:uplift_release} show the distribution of the uplift reasons, as well as the distribution of fault-inducing uplifts and clean uplifts for each reason.
We observe that, in the Beta channel, most patches are uplifted because of a wrong functionality, crash, security vulnerability, incompatibility with some major websites, or to introduce/remove a feature. Most regressions are introduced by the uplifts that resolved wrong functionalities, crash, and security issues.
For some uplift reasons, including improvement, resolving add-on/plug-in incompatibility and compiling errors, few patches lead to faults in our studied sample. However, a high percentage of patches resolving performance and rendering problems introduced new regressions.

In the Release channel, we observe the same top five uplift reasons. Compared to the Beta channel, there are fewer regressions; implying that these uplifted patches may have been more carefully scrutinized, the rules for approval on the Release channel being more strict. 
The fault-inducing patches only concentrated on five uplift categories: crash, hang, security, performance degradation, and incorrect rendering. Especially, most patches for incorrect rendering lead to future faults. These results suggest that, although developers prudently uplift patches in the Release channel, they still need to carefully review patches belonging to the aforementioned categories in order to prevent delivering faults to users.

Through the interview, we learn that release managers take into account several factors when deciding whether to approve or reject a patch uplift request. 

\begin{enumerate}
	\item Importance of the issue. This is measured through the impact that rejecting the uplift would have on users. 
	\item Risk associated with the patch. Release managers share the same view on the risks. They generally trust developers' words, unless they have had bad experiences with them (\eg{} developers who caused regressions and did not fix them); they evaluate the risk of the patch by looking at its size and complexity, the presence/absence of automated tests, the reviewers of the patch. In case of doubts, release managers consult other release managers or engineering managers to get a clearer picture.
	\item Timing of the uplift in the Aurora/Beta cycle. They tend to trust more patches that have been in Nightly for some time and patches that are far from the next release date. They almost always accept uplifts requested during the first weeks of the Aurora cycle.
	\item Verification of the patch. In particular for more stable channels, they make sure that the patch has been verified to actually fix the problems it was supposed to fix. If needed, they ask QA to manually verify the patch. If it is a patch that fixes a Nightly crash,  before uplifting the patch to Aurora, they will verify if users are no longer reporting the crash.
\end{enumerate}
They remarked that the uplift bar gets higher as they are getting closer to release. After the middle point of the Beta cycle, they only accept patches fixing high security issues, high-volume crashes, severe recent regressions, severe performance issues or memory leaks.

We presented the release managers with the results of our quantitative and qualitative analysis and collected the following observations.\\
\textbf{They found that the response delta information is interesting}. After thinking about it, they all gave us similar replies. When they are evaluating a complex issue and are still undecided, they will not make the call immediately. One release manager said that \emph{``when I reject something, I won't make the call immediately. I will think about it before doing it, in case I change my mind or new facts are coming in the equation"}.\\
\textbf{Regarding the landing delta}, they were surprised, as they thought they were more likely to accept patches with a higher landing delta (that is, patches that have been in Nightly for longer). They have also said that they are almost always accepting patches during the first four weeks of the Aurora cycle, which would explain this discrepancy (as those patches have a small landing delta).

The interviewed release managers also told us that they take into account the fault-proneness of components when making uplift decisions; which is in line with what we found (some components have a smaller acceptance rate). One release manager told us that \emph{``some components always come out as causing the most regressions, \eg{} graphics layers, DOM".} Regarding the trust in developers, they all mentioned the assessment of risk as one of the first factors. One release manager explained that \emph{``when they seem really overconfident or aren't telling me the whole story I lose some trust"}, another one stated that \emph{``some developers are taking a lot of risks, some other less and are super reactive to fix potential fallout"}. This finding is consistent with the uplift criteria followed at Facebook~\cite{rossi_talk}, where release managers tend to trust developers who introduced less regressions in the past.

Regarding uplift reasons, release managers were not surprised that test and compile changes are less frequent than others. They argued that these kinds of changes are really hard to move from the Nightly channel to a stabilization channel (build or test failures, unless they happen on really particular configurations, are noticed as soon as a patch is applied, since tests are run for every changeset). For the same reasons, they were not surprised that the uplift regressions are rarely compile-related.

Release managers argued that the information about the distribution of uplift reasons is useful for their future decision-making. They were initially surprised to see that crash and security-related uplifts often caused regressions, but they thought that the urgency of those fixes might degrade their quality. They were also interested in the results regarding the categories where a high proportion of uplift patches caused regressions (\eg{} performance uplifts). They said that they will start to take this information into account when deciding about uplifts, and will be more careful with the uplifts in those categories.  
\subsection*{RQ2: \RQtwo}
\noindent
\emph{\textbf{Motivation.}}
In Firefox' Aurora, Beta and Release channels, we found respectively 8.8\%, 8.3\%, and 7.9\% of uplifted patches that introduced regressions in the system. These patches not only decrease the users-perceived software quality, but also increase development costs, since developers, testers and release managers have to rework the faulty patches. 
In \textbf{RQ1}, we have identified some characteristics of patches that are taken into account by Mozilla release managers during patch uplifts. In this research question, we are interested in identifying the characteristics of uplifted patches that introduced faults in the system. 

\medbreak
\paragraph{1) Quantitative Analysis}

\emph{\textbf{Approach.}}
We apply the SZZ algorithm (described in Section \ref{sec:szz}) on all fault-fixing changes to identify uplifted patches that introduced a fault in the system. Next, we classify the uplifted patches into two groups: fault-inducing uplifts and clean uplifts. We use the 22 metrics listed in Table \ref{tab:all_metrics} to assess the differences between these two groups. For each ($m_i$) metric, we test the following hypothesis:

$H_i^{02}$: \emph{there is no difference between the values of $m_i$ for uplifted patches that introduced a fault in the system and those that did not.}

Similar to \textbf{RQ1}, we use the Mann-Whitney U test and Cliff's Delta effect size to accept or reject the hypotheses, and assess the magnitude of the differences between fault-inducing uplifts and clean uplifts. We also test the hypotheses for all three channels.


\begin{table}[t]
\centering\scriptsize
 \caption{Fault-inducing Uplifts vs. Clean uplifts.}
 \label{tab:rq2_res}
 \begin{tabular}{lp{1.7cm}rrrr}
	\toprule
    Channel & Metric & Faulty & Clean & $p$-value & Effect size \\
    \midrule
    \emph{Aurora}
    & Patch size & 155.0 & 34.0 & \textbf{5.59e-65} & large \\
    & Prior changes & 362.5 & 164.0 & \textbf{3.80e-10} & small \\
    & LOC & 903.6 & 457.4 & \textbf{2.23e-06} & small \\
    & Cyclomatic & 2.5 & 2.0 & \textbf{1.08e-06} & small \\
    & \# of functions & 34.3 & 17.0 & \textbf{2.25e-06} & small \\
    & Max. nesting & 2.7 & 2.0 & \textbf{5.14e-04} & negligible \\
    & Comment ratio & 0.2 & 0.1 & \textbf{4.00e-15} & small \\
    & Module number & 2.0 & 1.0 & \textbf{2.99e-24} & small \\
    & Closeness & 1.5 & 1.2 & \textbf{2.78e-13} & small \\
    & Betweenness & 45,221.9 & 880.7 & \textbf{2.65e-14} & small \\
    & PageRank & 1.7 & 1.4 & \textbf{1.95e-15} & small \\
    & \# of comments & 26.0 & 20.0 & \textbf{1.76e-09} & small \\
    & Developer exp. & 28.5 & 10.0 & \textbf{1.19e-18} & small \\
    & Reviewer exp. & 9.0 & 2.0 & \textbf{6.63e-09} & small \\
    & Comment words & 10.0 & 2.0 & \textbf{9.08e-07} & small \\
    & Developer senti. & -3 & -3 & \textbf{8.92e-04} & negligible \\
    & Owner sentiment & -2 & -1 & \textbf{1.66e-04} & negligible \\
    \hline
    \emph{Beta}
    & Patch size & 141.0 & 32.0 & \textbf{6.44e-33} & large \\
    & Prior changes & 268.0 & 156.5 & \textbf{1.02e-03} & small \\
    & LOC & 895.5 & 476.3 & \textbf{1.66e-03} & small \\
    & Cyclomatic & 2.5 & 2.0 & \textbf{3.69e-03} & small \\
    & \# of functions & 37.0 & 18.0 & \textbf{3.13e-03} & small \\
    & Max. nesting & 2.7 & 2.2 & \textbf{0.01} & negligible \\
    & Comment ratio & 0.2 & 0.1 & \textbf{4.61e-05} & small \\
    & Module number & 2.0 & 1.0 & \textbf{7.45e-12} & small \\
    & Closeness & 1.6 & 1.2 & \textbf{2.87e-07} & small \\
    & Betweenness & 35,661.7 & 1,327.8 & \textbf{6.00e-08} & small \\
    & PageRank & 1.7 & 1.4 & \textbf{1.08e-06} & small \\
    & \# of comments & 28.0 & 22.0 & \textbf{1.18e-04} & small \\
    & Comment words & 8.0 & 3.0 & \textbf{0.04} & negligible \\
    & Developer exp. & 29.0 & 10.0 & \textbf{1.33e-08} & small \\
    & Reviewer exp. & 10.0 & 2.0 & \textbf{3.35e-05} & small \\
    & Owner sentiment & -2 & -1 & \textbf{4.14e-03} & small \\
    \hline
    \emph{Release}
    & Patch size & 108.0 & 27.0 & \textbf{2.07e-03} & large \\
    \bottomrule
 \end{tabular}
 \vspace{-15pt}
\end{table}

\smallbreak
\emph{\textbf{Results.}}
Table \ref{tab:rq2_res} summarizes differences between the characteristics of uplifted patches that introduced a fault in the system and those that did not. We observe that fault-inducing uplifts \BLUE{have significantly larger patch size ($m_{11}$)} 
than clean ones, across all three channels. The effect size of the difference is large. This implies that patches with larger modifications are more likely to introduce a regression if uplifted. We observed the following on the different channels:
\begin{itemize}
\item On Aurora and Beta channels, fault-inducing uplifts tend to have more complex code in terms of LOC, cyclomatic complexity, number of functions, and number of modules. These patches often contain classes that are connected to many other classes, in terms of closeness, betweenness and PageRank. Fault-inducing uplifts also tend to have higher comment ratios and tend to change files that were changed more frequently. 
    Interestingly, fault-inducing uplifts are frequently submitted by developers or reviewers with high experience. 
    Fault-inducing uplifts also have a larger amount of comments than clean uplifts. A large number of comments may be a sign that developers are struggling with the patch, which may explain the high fault-proneness.
Although fault-inducing uplifts and clean uplifts also display other significant differences (as shown in Table \ref{tab:rq2_res}), the magnitude of these differences is negligible.
\item For the Release channel, we do not observe a significant difference between fault-inducing uplifts and clean uplifts for the above metrics. 
\end{itemize}

\textbf{Overall, we reject $H_{11}^{02}$, \ie{} fault-inducing uplifts \BLUE{have larger patch size}
than clean uplifts. Release managers should pay attention to \BLUE{large} patches and reviewers should scrutinize them carefully. 
Although the effect of other characteristics is channel dependent, in Aurora and Beta, we observe that patches with high complexity and centrality tend to lead to faults. Uplift requests submitted by experienced developers and reviewers also tend to lead to regressions.} 

Similar to \textbf{RQ1}, we examined patch uplifts per component, and observed that patch uplifts affecting certain components (\eg{} \textit{Graphics} component) are more likely to cause regressions than others. 
Some of the components with the highest fault-inducing rates also have a low approval rate; probably because the release managers were acting based on their previous experiences with those components (for example, the \textit{Web Audio} component). Components like the \textit{Audio/Video}, which are involved in multiple patch uplift operations, also have the highest fault-inducing rates; these components would be inherently more prone to faults because of their complexity, or technical debt.

We made a similar observation regarding developers' submitting uplift requests. 
Many developers who submitted multiple uplift requests appear in the list of developers with high fault-inducing rates; perhaps, by uplifting more patches, they are taking more risks.


\medbreak
\paragraph{2) Qualitative Analysis}
To understand the root cause of faults in uplifted patches, we conduct a qualitative study.

\emph{\textbf{Approach.}}
We manually examined uplifted patches (from the samples selected in \textbf{RQ1}) that introduced faults, and classified the reasons behind the faults. 
Inspired by the work of Tan et al~\cite{tan2014bug}, we defined seven possible root causes for uplift faults (as shown in Table \ref{tab:fault_reasons}). We identified respectively 132 and 17 fault-inducing uplifts from the Beta and Release samples chosen in \textbf{RQ1}, 
and performed a card sorting to classify each of the faults into one or multiple causes. 
As in \textbf{RQ1}, the first and the second authors individually read the issue reports and their fault-fixing patches to understand the root causes of the faults (\ie{} the reason why their corresponding uplifted patches caused the faults) and classified these root causes along our seven categories. Similar to \textbf{RQ1}, disagreements were resolved through discussions. 

We also interviewed release managers, asking them the following question:
\emph{What are the characteristics of fault-inducing patches that you are not currently taking enough into account but could be considered in the future?}

\begin{table}[t]
	\centering
	\scriptsize
	\caption{Fault reasons and descriptions.}
	\begin{tabular}{ | >{\RaggedRight} p{2cm} | p{6cm} | }
		\hline
		\textbf{Reason} & \textbf{Description} \\ \hline
		Memory & Memory errors, including memory leak, overflow, null pointer dereference, dangling pointer, double free, uninitialized memory read, and incorrect memory allocation.\\ \hline
		Semantic & Semantic errors, including incorrect control flow, missing functionality, missing cases of a functionality, missing feature, incorrect exception handling, and incorrect processing of equations and expressions.\\ \hline
		Third-party & Errors due to incompatibility of drivers, plug-ins or add-ons.\\ \hline
		Concurrency & Synchronization problems between multiple threads or processes, \eg{} incorrect mutex usage.\\  \hline
		Compile & Compile-time errors. \\ \hline
		Other & Other errors.\\ \hline	
	\end{tabular}
	\label{tab:fault_reasons}
\vspace{-15pt}
\end{table}

\begin{figure}[t]
\centering
\includegraphics[width = \columnwidth]{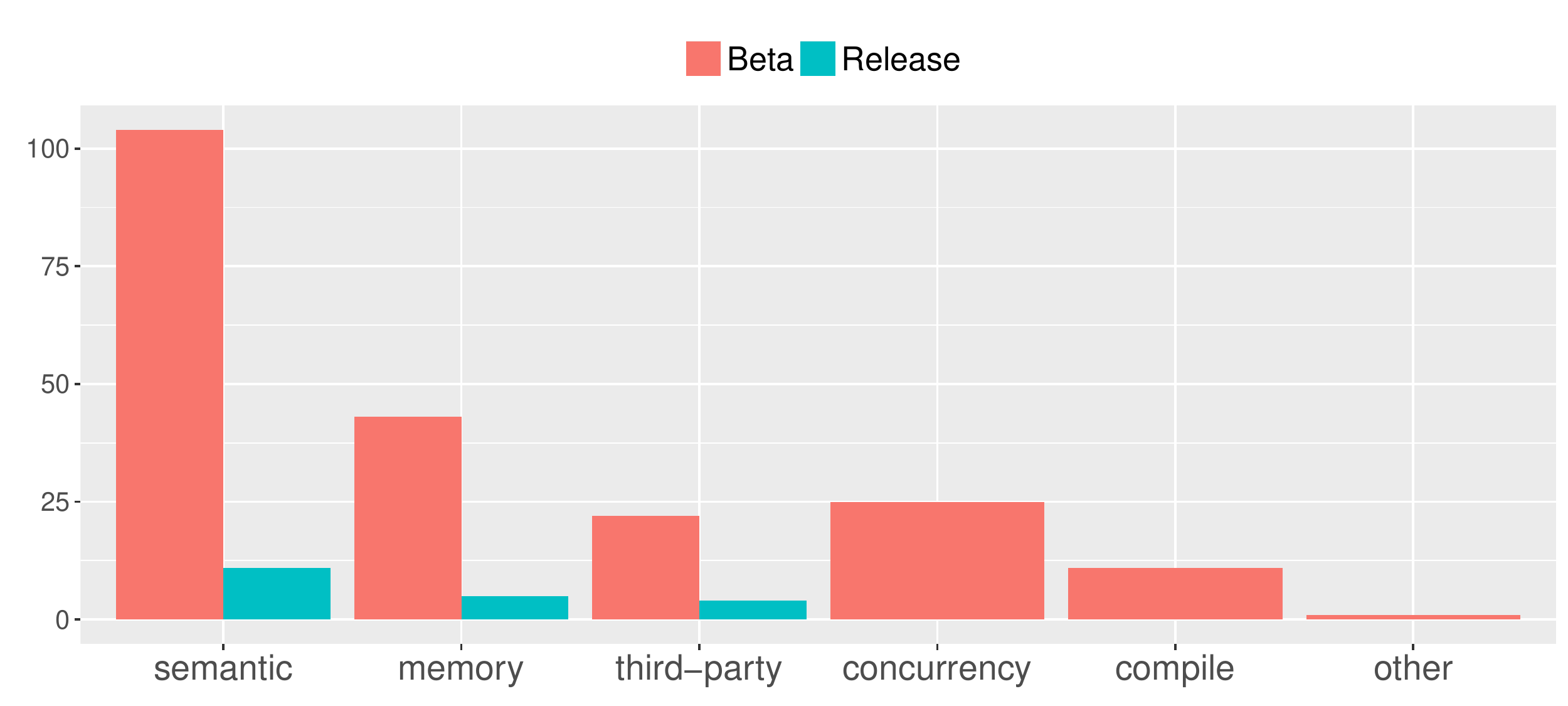}
\caption{Reasons of fault-inducing uplifts.}
\label{fig:fault_reason}
\vspace{-10pt}
\end{figure}


\smallbreak
\emph{\textbf{Results.}}
Figure \ref{fig:fault_reason} depicts the distribution of the reasons why fault-inducing uplift introduced regressions. In both channels, semantic and memory-related errors are dominant root causes of the uplift regressions. With a detailed check on the patches, we find that many memory errors are due to null pointer dereference and memory leak.
In addition, incompatibility of plug-ins and drivers also cause uplift regressions in both channels. Concurrency issues are ranked as a popular cause for Beta's uplift regressions, but we do not find any \BLUE{example} of this category in the Release channel.
In general, our results suggest that, when uplifting a patch, \textbf{release managers need to carefully check for potential faults on the program's semantic meaning, memory operations, synchronization, and third-party extension's compatibility}.

In the interview,
\textbf{all the release managers agreed that it would be beneficial for them to have more detailed information about the complexity of the patches they are asked to evaluate and more information about the history of the components involved in these patches}. This resonates with our findings. Release managers were surprised to see that fault-inducing patches were more likely to be written by more experienced developers and reviewed by more experienced reviewers. They guess that these developers/reviewers are assigned to more complex tasks with more complex solutions. A release manager told us that \emph{``if you call in the big guns, then it's a warning sign"}.


The fault categorization was also interesting for the release managers, who told us that Mozilla is about to employ more static analysis tools (\eg{} Coverity~\cite{coverity}) and to move some of their code from C++ to a safer language (\eg{} Rust).
It is promising for them to see how many memory and concurrency faults can be avoided by using these techniques, and how many semantic and third-party faults can be reduced by enhancing code review or testing efforts.

\section{Discussion}\label{sec:discussion}

According to the results of \textbf{RQ1} and \textbf{RQ2}, there are statistically significant differences between the characteristics of uplifted patches that introduced regressions and those that were integrated successfully (\ie{} clean uplifts that did not induce faults). Also, fault-inducing uplifts are in the majority of cases uplifts that were meant to resolve wrong functionalities, crashes, security vulnerabilities, and incompatibilities with websites. Furthermore, incorrect semantic code and memory operations are the most important root causes of uplift regressions.

We believe that release management teams could leverage these findings to build classifiers capable of automatically assessing the risk associated with patch uplift candidates and recommend patches that can be uplifted safely.

Exploring the possibility of building such classifiers is part of our future work agenda. 
\section{Threats to Validity}
\label{sec:threats}


\emph{Construct validity} threats are concerned with the relationship between theory and observation. 
Previous studies~\cite{kamei2013large,kim2011crashes} suggested that complex code is a good indicator of fault-proneness. We confirm this point in this study. However, we found that fault-inducing patches are more likely to be submitted by experienced developers, which contradicted our expectations. We attribute this outcome to the fact that experienced developers are often assigned to difficult issues, whose resolution tend to be more complex. Also, release managers might overlook risks associated to patches submitted by experienced developers, as these developers are often more trusted than others.

\emph{Internal validity} threats concern factors that may affect a dependent variable and were not considered in the study. We paid attention not to violate the assumptions of the statistical tests that are performed in the paper. Specifically, in \textbf{RQ1} and \textbf{RQ2}, we applied non-parametric tests that do not require making assumptions on the distribution of our dataset.

\emph{Conclusion validity} threats concern the relationship between the treatments and the outcome. Before conducting the case study, we limited our studied dataset within a duration that covers consecutive series of relatively stable periods on all the three uplift channels.
In addition, we used a keyword matching heuristic to identify fault-related issues. We manually validated a random sample of 380 issues. All the authors of this paper participated in the validation. Whenever there were diverging opinions, we set up a meeting and discussed the issue until a consensus was reached. As a result, we found that our heuristic can achieve a precision of 87.3\% and a recall of 78.2\%, when identifying fault-related issues.
Moreover, we performed a manual classification of the uplift reasons and the root causes of uplift regressions. To mitigate potential bias that may result from our subjective opinions, we also discussed on each of our classification conflicts until reaching a consensus. However, as any other taxonomic study, we cannot guarantee a 100\% of accuracy on our classification results. Future replications are welcomed to validate our work.
Another issue on the manual classification is that, although we randomly chose our samples by applying a confidence level of 95\% and a confidence interval of 5\%, our samples might not precisely reflect the distributions of the uplift reasons and--or root causes of uplift regressions on the whole Firefox dataset. Further investigations on larger data sets are desirable. 

\emph{External validity} threats are concerned with the generalizability of our results. In this paper, we only studied Mozilla Firefox. First, Mozilla Firefox is the most studied system for issues related to rapid releases; moreover, the system's data are publicly available.  We also have the opportunity to perform both quantitative and qualitative analyses (including the interviews with release managers) on this system.
However, we should recognize that our findings may not be generalizable to other systems.
In the future, we plan to collaborate with other software organizations, to validate and extend the results of this work. In addition, more studies on other systems with other programming languages are suitable to further validate our results. To facilitate future replication studies, we share our datasets and scripts at: \url{https://github.com/swatlab/uplift-analysis}.

\section{Related Work}
\label{sec:related_work}
Patch uplift is an activity performed during the release engineering process. Hence, in this section, we present and discuss relevant literature on release engineering. 

Release engineering encompasses all the activities aimed at ``building a pipeline that transforms source code into an integrated, compiled, packaged, tested, and signed product that is ready for release''~\cite{adams2015practice}.

Since the adoption of the rapid release model~\cite{khomh2012faster} by Mozilla in 2011, a plethora of studies have focused on the impact of rapid release strategies on software quality. Khomh et al.~\cite{khomh2012faster} compared crash rates, median uptime, and the proportion of post-release bugs between the versions of Firefox that followed a traditional release cycle and those that followed a rapid release cycle. They observed that short release cycles do not induce significantly more bugs. However, compared to traditional releases, users experience bugs earlier during software execution. Nevertheless, they also observed that post-release bugs are fixed faster under the rapid release model. 
Da Costa et al.~\cite{costa2016msr} studied the impact of Mozilla's rapid release cycles on the integration delay of addressed issues. They found that, compared to the traditional release model, the rapid release model does not deliver addressed issues to end users more quickly, which is contrary to expectations.

Another important aspect of release engineering that has been investigated by the community is the integration of urgent patches that are used to fix severe problems, such as frequent crashes or security bugs, or to introduce important features. Urgent patches break the balance between new feature work and software quality, and hence could lead to faults and failures. Hassan et al.~\cite{hassan2016empirical} investigated emergency updates for top Android apps and identified eight patterns along the following two categories: ``updates due to deployment issues'' and ``updates due to source code changes''. They suggested to limit the number of emergency updates that fall in these patterns, since they are likely to have a negative impact on 
users' satisfaction. 
In a recent work, Lin et al.~\cite{linstudying} empirically analyzed urgent updates in 50 most popular games on the Steam platform, and 
observed that the choice of the release strategy affects the proportion of urgent updates, \ie{} games that followed a rapid release model had a higher proportion of urgent patches in comparison to those that followed the traditional release model.
Rahman et al.~\cite{rahman2015release} examined the ``rush to release'' period on Linux and Chrome. They observed that experienced developers are often allowed to make changes right before stabilization occurs and these changes are added directly to the stabilization line. They also found that there is a rush in the number of commits right before a new release is added to the stabilization channel, to add final features. In a following work, Rahman et al.~\cite{rahman2016feature} observed that feature toggles~\cite{feature_toggle} can effectively turned off faulty urgent patches, which limits the impact of faulty patches.

To the best of the authors' knowledge, none of these prior works has empirically investigated how urgent patches in the rapid release model affect software quality in terms of fault-proneness, and how the reliability of the integration of urgent updates could be improved. This paper fills this gap in the literature by investigating the reliability of the Mozilla's uplift process, since uplifted patches are urgent updates.


\section{Conclusion}
\label{sec:conclusion}
Mozilla follows a rapid release model, which uses 18 weeks to deliver fault fixes and new features to users. Frequently, certain patches that fix critical issues, or implement high-value features are promoted directly from the development channel to a stabilization channel, because they are too urgent and cannot wait for the next release train. This practice, known as \emph{patch uplift}, is risky because the time allowed for the stabilization of the uplifted patches is short. In average, 8\% of uplifted patches introduced a regression in the code of Firefox. 
%
In this paper, we investigated the decision making process of patch uplift at Mozilla and observed that 
release managers are more inclined to accept patch uplift requests that concern certain specific components, and--or that are submitted by certain specific developers. We examined the characteristics of uplifted patches that introduced regressions in the code and found that they are more complex than clean uplifts, and they tend to \BLUE{change a higher number of lines of code}. Most regressions are caused by patch uplifts aimed at fixing wrong functionalities and crashes. The most common root causes of faults in uplifted patches are semantic and memory errors.
Reviewers and release managers should carefully inspect complex patches before allowing their uplift. 


\balance
\bibliographystyle{IEEEtran}
\bibliography{anle,marco,__reading,PURE,CASCON2009,foutsekh}
\end{document}